\documentclass[12pt]{article}

  \usepackage{graphicx}
  \usepackage{epsfig}
  \usepackage{amssymb}
  \usepackage{subfigure}
  \usepackage{axodraw}
\usepackage{hyperref}
  \usepackage{amsmath} 
\usepackage{float}
\usepackage{cite}
\usepackage{color}
\evensidemargin - .5truecm
\allowdisplaybreaks[4]

\begin{document}

\newcommand{\be}{\begin{equation}}
\newcommand{\ee}{\end{equation}}
\newcommand{\nn}{\nonumber}
\newcommand{\bea}{\begin{eqnarray}}
\newcommand{\eea}{\end{eqnarray}}
\newcommand{\bfig}{\begin{figure}}
\newcommand{\efig}{\end{figure}}
\newcommand{\bc}{\begin{center}}
\newcommand{\ec}{\end{center}}
\newcommand{\tnovez}{{\mathcal T}_{9}^{(0)}}
\newcommand{\tnove}{{\mathcal T}_{9}}

\newenvironment{appendletterA}
{
  \typeout{ Starting Appendix \thesection }
  \setcounter{section}{0}
  \setcounter{equation}{0}
  \renewcommand{\theequation}{A\arabic{equation}}
 }{
  \typeout{Appendix done}
 }
\newenvironment{appendletterB}
 {
  \typeout{ Starting Appendix \thesection }
  \setcounter{equation}{0}
  \renewcommand{\theequation}{B\arabic{equation}}
 }{
  \typeout{Appendix done}
 }

%
%
%
%

\begin{titlepage}
\nopagebreak
{\flushright{
        \begin{minipage}{5cm}
         IFT-UAM/CSIC-18-118
         HU-EP-18/37
        \end{minipage}        }

}
\renewcommand{\thefootnote}{\fnsymbol{footnote}}
\vspace*{1.2cm}   
\begin{center}
\boldmath
{\Large\bf A Numerical Routine for the \\[7pt] Crossed Vertex Diagram 
with a Massive-Particle Loop}\unboldmath
\vskip 1.cm
{\large  Roberto Bonciani\footnote{Email:
roberto.bonciani@roma1.infn.it}},
\vskip .2cm
 {\it Dipartimento di Fisica, Sapienza - Universit\`a di Roma, 00185, Rome, Italy \\
and INFN Sezione di Roma, 00185, Rome, Italy}
\vskip .2cm
{\large Giuseppe Degrassi\footnote{Email: degrassi@fis.uniroma3.it}},
\vskip .2cm
{\it Dipartimento di Matematica e Fisica, Universit{\`a} di Roma Tre,
 00146 Rome, Italy\\
 and INFN, Sezione di Roma Tre, 00146 Rome, Italy} 
\vskip .2cm
{\large Pier Paolo Giardino\footnote{Email: pier.giardino@uam.es}}
\vskip .2cm
{\it Instituto de F\'isica Te\'orica UAM/CSIC, Universidad Aut\'onoma de Madrid, 28049, Madrid, Spain} 
\vskip .2cm
{\large Ramona Gr\"{o}ber\footnote{Email: ramona.groeber@physik.hu-berlin.de}}
\vskip .2cm
{\it Institut f\"ur Physik, Humboldt-Universit\"at zu Berlin, 12489 Berlin, Germany \\ 
and Institute for Particle Physics Phenomenology, Department of Physics, Durham University, Durham, DH1 3LE, UK} 
\end{center}
\vskip 0.7cm

\begin{abstract}
We present an evaluation of the two master integrals for the crossed vertex
diagram with  a closed loop of top quarks that allows for an easy numerical
implementation. The differential equations obeyed by the master 
integrals are used to generate power series expansions centered around all 
the singular points. The different series are then matched numerically with 
high accuracy in intermediate points. The expansions allow a fast and precise 
numerical calculation of the two master integrals in all the regions of the 
phase space. A numerical routine that implements these expansions is presented.


\end{abstract}
\vfill
\end{titlepage}    

\setcounter{footnote}{0}

\section{Introduction}

In the last years, we witnessed an impressive progress in the analytic
calculation of multi-loop Feynman diagrams. This progress was mainly due
to a procedure which is by now standard and consists in the
reduction of the dimensionally regularized scalar integrals to the
Master Integrals (MIs)
\cite{Tkachov:1981wb,Chetyrkin:1981qh,Laporta:2001dd,Gehrmann:1999as,Anastasiou:2004vj,Lee:2012cn,Smirnov:2008iw,Studerus:2009ye,vonManteuffel:2012np,Smirnov:2013dia,Smirnov:2014hma,Lee:2013mka,Maierhoefer:2017hyi},
and their calculation using the differential equations \cite{Kotikov:1990kg,Remiddi:1997ny,Gehrmann:1999as,Argeri:2007up,Henn:2014qga}. 

With this procedure, it was possible to calculate massless quantum
corrections to important processes in collider physics, that are now
known to three and four loops
\cite{Moch:2005tm,Gehrmann:2006wg,Heinrich:2007at,Heinrich:2009be,Lee:2010cga,Henn:2013tua,Anastasiou:2015ema,vonManteuffel:2016xki,Lee:2017mip,Lee:2016ixa}. These
corrections can be usually expressed in terms of generalized
polylogarithms (GPLs)
\cite{Goncharov:polylog,Goncharov2007,Remiddi:1999ew,Vollinga:2004sn}.
While sometimes higher-order massive corrections can also be expressed
in terms of GPLs
\cite{Henn:2016kjz,Henn:2016tyf,Lee:2018nxa,Lee:2018rgs,Chen:2018fwb},
they reveal in general a more complicated structure. This is for
instance the case of the two MIs of the equal-mass two-loop
sunrise. The related system of first-order linear differential
equations cannot be decoupled and it admits solutions in terms of
complete elliptic integrals of the first and second kind
\cite{Laporta:2004rb,Adams:2013kgc,Remiddi:2016gno,Bloch:2013tra,Bloch:2014qca,Adams:2015ydq,Bogner:2017vim}. This
is also the case of three-
\cite{Aglietti:2007as,vonManteuffel:2017hms} and four-point functions
\cite{Bonciani:2016qxi,Adams:2018bsn,Adams:2018kez} that were
considered recently and whose solutions are expressed as iterated
integrals over elliptic kernels multiplied by polylogarithmic
terms. The study of these new functions has just started
\cite{Ablinger:2017bjx,Broedel:2017kkb,Broedel:2017siw,Remiddi:2017har,Broedel:2018iwv}.

In this article, we focus on the calculation of the two MIs of the vertex 
crossed topology with a closed
heavy-quark loop. These two MIs were studied in detail in
Ref.~\cite{vonManteuffel:2017hms},
where the authors worked out completely their solution in terms of
repeated integrations over elliptic kernels.
They enter the calculation of several processes
at the two-loop level in perturbation theory, as the production of
top-antitop pairs in hadronic collisions
\cite{Bonciani:2008az,Bonciani:2009nb,Bonciani:2010mn,
  vonManteuffel:2013uoa,Bonciani:2013ywa,Adams:2018bsn,Adams:2018kez,
  Mastrolia:2017pfy,DiVita:2018nnh}, di-photon or di-jet production
\cite{Becchetti:2017abb} and they are part of the coefficients of the
$p_T$ expansion of the double Higgs production cross section, as
discussed in Ref.~\cite{Bonciani:2018omm}.

Our goal is to present a Fortran numerical routine that can be easily
used to evaluate the MIs for every real value of the dimensionless
parameter $x=-S/m^2$, which the MIs depend on, with double precision.
The approach  we use is a semi-analytical approach to the solution
of the differential equations, namely the expansion of the
differential equation near singular points.
It was proposed in Ref.~\cite{Pozzorini:2005ff} for the sunrise with three
equal masses.  In Ref.~\cite{Aglietti:2007as} the method was applied to a
three-point function\footnote{See Refs.\cite{Lee:2017qql,Lee:2018ojn} for
  two recent publications on the method.} occurring in the calculation
of the MIs that are involved in the two-loop corrections to the
electroweak form factor \cite{Aglietti:2003yc,Aglietti:2004tq}.

The paper is structured as follows. In Section~\ref{sec2}, we
discuss the MIs entering the 6-denominator topology of Fig.~\ref{fig1}.
We focus on the two crossed MIs (${\mathcal T}_{9},\: {\mathcal T}_{10}$)
for which  we present the relevant second-order linear differential
equation that will be solved expanding the solution by series near the
singular points.  Section ~\ref{sec3} is devoted to the discussion of the
solution for $\tnove$ in the region $x \geq 0$. We present first the series in
the two singular points $x=0$ and  $x=16$ and their matching. Then,
we  discuss the expansion at infinity and how it can be matched to the
expansion at $x=16$. In Section~\ref{sec4}, we present the solution
for $\tnove$ in the region $x < 0$ obtained via the analytic
continuation in the high-energy time-like region.  In Section~\ref{SecMI}, we
discuss the evaluation of the second master
integral. Finally, in Section~\ref{fortranroutine}, we present the
Fortran routine.

\section{The Differential Equations for the two crossed 
  Master Integrals}
\label{sec2}

We consider a process in which two massless particles with incoming
momenta $p_1$ and $p_2$, such that $p_1^2=p_2^2=0$, annihilate into a
particle with momentum $p=p_1+p_2$. We define the Mandelstam invariant
$S=-(p_1+p_2)^2 $ and the dimensionless parameter
\be 
x = - \frac{S}{m^2} = - s \, ,
\ee
where $s=S/m^2$ and $m$ is the mass of a massive state that runs into the loops.
\begin{figure}
\vspace*{10mm}
\centering
\begin{picture}(0,0)(0,0)
\SetScale{0.8}
\SetWidth{1}
\Line(-80,70)(-60,58)
\Line(0,23)(40,0)
\Line(40,0)(0,-23)
\Line(-60,-58)(-80,-70)
\DashLine(40,0)(60,0){2}
\SetWidth{4}
\Line(-60,58)(0,23)
\Line(0,-23)(-60,-58)
\Line(-60,58)(0,-23) 
\Line(-60,-58)(0,23) 
\end{picture}
\vspace*{15mm}
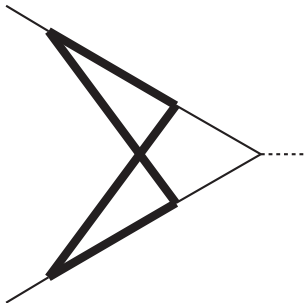
\caption[]{The 6-denominator topology. Internal plain thin lines represent
  massless propagators, while thick lines represent the massive propagator.
  External plain thin lines represent massless particles on their mass-shell.}
\label{fig1}
\efig

The 6-denominator topology we are interested in relevant for  this process 
is shown in Fig.~\ref{fig1}.  The dimensionally regularized scalar integrals
belonging to that topology can be expressed in terms of
\be
\int {\mathcal D}^dk_1 {\mathcal D}^dk_2
\frac{1}{D_1^{a_1}D_2^{a_2}D_3^{a_3}D_4^{a_4}D_5^{a_5}D_6^{a_6}D_7^{a_7}}
\, .
\label{fam}
\ee
In Eq.~(\ref{fam}), $D_i$, $i=1,...,7$, are the denominators
to which the following routing is assigned
\bea
D_{1,..7} &=& \Bigl\{ k_1^2+m^2, (p_1+k_1)^2+m^2, k_2^2+m^2, (p_2+k_2)^2+m^2, 
(p_1-k_1-k_2)^2,\nn\\
& &  (p_2+k_1+k_2)^2, (k_1+k_2)^2 \Bigr\} \, ,
\eea
with $k_1$ and $k_2$  the loop momenta;
$a_i$, with $i=1,...,7$, are integer numbers,
$d=4-2\epsilon$ is the dimension of the space-time, and the
normalization is such that\footnote{Note that we present, in the paper and in the routine, the 
euclidean version of the MIs, before Wick rotation.
} 
\be 
{\mathcal D}^dk_i = \frac{d^d k_i}{4 \pi^{\frac{d}{2}}\Gamma(1+\epsilon)} \left( \frac{m^2}{\mu^2}
\right) ^{\epsilon} \, , 
\label{normalization}
\ee 
where $\mu$ is the scale of dimensional regularization.

The reduction to the MIs of the family in Eq.~(\ref{fam}) are
performed using the computer programs
\texttt{FIRE} \cite{Smirnov:2008iw,Smirnov:2013dia,Smirnov:2014hma}
and \texttt{Reduze 2} \cite{Studerus:2009ye,vonManteuffel:2012np}.
There are 10 MIs in total, shown in Fig.~\ref{mis}.
All of them are known in the literature from previous works 
\cite{Bonciani:2003te,Bonciani:2003hc,Aglietti:2004tq,vonManteuffel:2017hms}. 

We focus on the evaluation of ${\mathcal T}_{9}$ and ${\mathcal T}_{10}$
using the semi-analytic approach followed in
Refs.~\cite{Pozzorini:2005ff,Aglietti:2004ki}.
We concentrate on the system of first-order linear differential
equations that involves the two coupled 6-denominator MIs ${\mathcal
  T}_{9}$ and ${\mathcal T}_{10}$. The two MIs are finite in
$\epsilon$. Moreover, in all the processes mentioned in the
introduction, at the NNLO, they enter in the calculation of the finite
part of the cross section so that only the ${\mathcal O}(\epsilon^0)$ 
is needed.
At the ${\mathcal O}(\epsilon^0)$, we find:
\bea \frac{d
  {\mathcal T}_{9}}{dx} &=& - \frac{2}{x} {\mathcal T}_{9} +
\frac{4m^2}{x} {\mathcal T}_{10} \, ,
\label{eq1} \\
\frac{d{\mathcal T}_{10}}{dx} &=& - \frac{1}{16m^2} \left( \frac{1}{x} -
\frac{1}{x-16} \right)  {\mathcal T}_{9} -  \left( \frac{1}{x} + \frac{1}{x-16}
\right) {\mathcal T}_{10} + \Omega_2(x) \, ,
\label{eq2} 
\eea
where $\Omega_2(x)$ contains the MIs of the subtopologies and, at this order in
$\epsilon$, is a function that can be expressed in terms of logarithms and
dilogarithms of the variable $x$. 
%
\bfig
\bc
\[ \vcenter{
\hbox{
  \begin{picture}(0,0)(0,0)
\SetScale{0.4}
  \SetWidth{1}
  \SetWidth{4}
\CArc(-15,0)(15,0,180)
\CArc(-15,0)(15,180,360)
\CArc(15,0)(15,180,360)
\CArc(15,0)(15,0,180)
\Text(0,-28)[c]{\footnotesize{(${\mathcal T}_{1}$)}}
\end{picture}}
}
\hspace{1.3cm}
\vcenter{
\hbox{
  \begin{picture}(0,0)(0,0)
\SetScale{0.4}
  \SetWidth{1.0}
\DashLine(-35,0)(-20,0){3}
\DashLine(20,0)(35,0){3}
\CArc(0,0)(20,0,180)
\CArc(0,0)(20,180,360)
  \SetWidth{4}
\CArc(60,0)(40,150,180)
\CArc(0,34.6)(40,300,330)
\CArc(30.20,17.60)(5.28,-34,153)
\Text(0,-28)[c]{\footnotesize{(${\mathcal T}_2$)}}
\end{picture}}
}
\hspace{1.3cm}
\vcenter{
\hbox{
  \begin{picture}(0,0)(0,0)
\SetScale{0.4}
  \SetWidth{1.0}
\DashLine(-35,0)(-20,0){3}
\DashLine(20,0)(35,0){3}
\Line(-20,0)(20,0)
  \SetWidth{4}
\CArc(0,0)(20,0,180)
\CArc(0,0)(20,180,360)
\Text(0,-28)[c]{\footnotesize{(${\mathcal T}_{3}$)}}
\end{picture}}
}
\hspace{1.3cm}
\vcenter{
\hbox{
  \begin{picture}(0,0)(0,0)
\SetScale{0.4}
  \SetWidth{1.0}
\DashLine(-35,0)(-20,0){3}
\DashLine(20,0)(35,0){3}
\CCirc(0,-20){5}{0.9}{0.9}
\Line(-20,0)(20,0)
  \SetWidth{4}
\CArc(0,0)(20,0,180)
\CArc(0,0)(20,180,360)
\Text(0,-28)[c]{\footnotesize{(${\mathcal T}_{4}$)}}
\end{picture}}
}
\hspace{1.3cm}
\vcenter{
\hbox{
  \begin{picture}(0,0)(0,0)
\SetScale{0.4}
  \SetWidth{1.0}
\DashLine(50,0)(35,0){3}
\Line(-10,30)(-25,30)
\Line(-10,-30)(-25,-30)
\Line(35,0)(-10,30)
\Line(35,0)(-10,-30)
\Line(-10,-30)(35,0)
  \SetWidth{4}
\CArc(-30,0)(35,305,55)
\Line(-10,30)(-10,-30)
\Text(0,-28)[c]{\footnotesize{(${\mathcal T}_{5}$)}}
\end{picture}}
}
\hspace{1.3cm}
\vcenter{
\hbox{
  \begin{picture}(0,0)(0,0)
\SetScale{0.4}
  \SetWidth{1.0}
\DashLine(50,0)(35,0){3}
\Line(-10,30)(-25,30)
\Line(-10,-30)(-25,-30)
\Line(35,0)(-10,30)
\Line(35,0)(-10,-30)
\Line(-10,-30)(35,0)
\CCirc(-10,0){4}{0.9}{0.9}
  \SetWidth{4}
\Line(-10,30)(-10,-30)
\CArc(-30,0)(35,305,55)
\Text(0,-28)[c]{\footnotesize{(${\mathcal T}_{6}$)}}
\end{picture}}
}
\hspace{1.3cm}
\vcenter{
\hbox{
  \begin{picture}(0,0)(0,0)
\SetScale{0.4}
  \SetWidth{1}
\DashLine(50,0)(35,0){3}
\Line(-10,30)(-25,30)
\Line(-10,-30)(-25,-30)
\CArc(25,34)(35,190,285)
  \SetWidth{4}
\Line(35,0)(-10,30)
\Line(-10,30)(-10,-30)
\Line(-10,-30)(35,0)
\Text(3,-28)[c]{\footnotesize{(${\mathcal T}_{7}$)}}
\end{picture}}
}
\hspace{1.3cm}
\vcenter{
\hbox{
  \begin{picture}(0,0)(0,0)
\SetScale{0.4}
  \SetWidth{1.0}
\DashLine(50,0)(35,0){3}
\Line(-10,30)(-25,30)
\Line(-10,-30)(-25,-30)
\Line(-10,-30)(35,0)
\Line(11,16)(35,0)
  \SetWidth{4}
\Line(-10,30)(11,16)
\Line(-10,-30)(11,16)
\Line(-10,30)(-10,-30)
\Text(0,-28)[c]{\footnotesize{(${\mathcal T}_{8}$)}}
\end{picture}}
}
\hspace{1.3cm}
\vcenter{
\hbox{
  \begin{picture}(0,0)(0,0)
\SetScale{0.4}
  \SetWidth{1.0}
\DashLine(50,0)(35,0){3}
\Line(-10,30)(-25,30)
\Line(-10,-30)(-25,-30)
\Line(11,-16)(35,0)
\Line(11,16)(35,0)
  \SetWidth{4}
\Line(-10,30)(11,16)
\Line(-10,-30)(11,-16)
\Line(11,-16)(-10,30)
\Line(-10,-30)(11,16)
\Text(0,-28)[c]{\footnotesize{(${\mathcal T}_{9}$)}}
\end{picture}}
}
\hspace{1.3cm}
\vcenter{
\hbox{
  \begin{picture}(0,0)(0,0)
\SetScale{0.4}
  \SetWidth{1.0}
\DashLine(50,0)(35,0){3}
\Line(-10,30)(-25,30)
\Line(-10,-30)(-25,-30)
\CCirc(-2,-14){5}{0.9}{0.9}
\Line(11,-16)(35,0)
\Line(11,16)(35,0)
  \SetWidth{4}
\Line(-10,30)(11,16)
\Line(-10,-30)(11,-16)
\Line(11,-16)(-10,30)
\Line(-10,-30)(11,16)
\Text(0,-28)[c]{\footnotesize{(${\mathcal T}_{10}$)}}
\end{picture}}
}
\]
\vspace*{7mm}
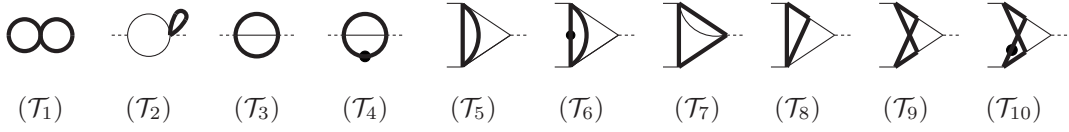
\caption{\label{mis} Master Integrals. The convention for the lines is as in
  Fig.~\ref{fig1}.
The dot represents a propagator raised to the second power.}
\ec
\efig
%

The system is equivalent to a single second-order linear differential
equation for one of the two MIs involved. Let us consider ${\mathcal  T}_{9}$.
We find:
\be
\frac{d^2 {\mathcal T}_{9}}{dx^2} + p (x)
\frac{d{\mathcal T}_{9}}{dx} + q(x) {\mathcal T}_{9} = \Omega(x),
\label{eqsec}
\ee
with ($\Omega(x)= (4 m^2/x) \Omega_2$)
\bea
p(x) &=& \frac{4}{x} + \frac{1}{x-16} \, , \\
q(x) &=&  \frac{9}{4x^2} - \frac{7}{64x} + \frac{7}{64(x-16)} \, , \\
\Omega(x) &=& \frac{1}{m^4} \Biggl\{ \frac{5}{64} \Biggl[ \frac{1}{256(x-16)} -
  \frac{1}{256 x}
- \frac{1}{16x^2} - \frac{1}{x^3} \Biggr] H(-r,-r,x) \nn\\
&& \hspace*{10mm} + \frac{3}{64}
\Biggl[\frac{1}{16(x-16)} - \frac{1}{16 x} - \frac{1}{x^2} \Biggr]
\frac{H(r,0,x)}{\sqrt{x(4-x)}} \Biggr\} \, ,
\label{eq10}
\eea
where we used the notation introduced in
Refs.~\cite{Aglietti:2004tq,Aglietti:2004nj}
for the repeated integration over square roots
\bea
H(-r,-r,x) & = & \int_0^x \frac{dt}{\sqrt{t(t+4)}} \int_0^t \frac{dt'}{\sqrt{t'(t'+4)}} \, , \\
H(r,0,x)   & = & \int_0^x \frac{dt}{\sqrt{t(4-t)}} \log{t} \, .
\eea
The function $H(-r,-r,x)$ is real when $x \geq 0$. In the Minkowski
region, $x \to -s-i 0^+$, with $s>0$, $H(-r,-r,x)$ is real
if $0<s<4$. For $s>4$, it develops an imaginary part due to the
branch cut of the square root.
The function $H(r,0,x)$ is real if $0<x<4$, while for $x>4$ the square
root of the integrand has a branch cut. The result is purely imaginary
and the sign depends on the sign of the small imaginary part that we
add to $x$ to chose on which part of the cut we are. The same happens
for the square root $\sqrt{x(4-x)}$ in Eq.~(\ref{eq10}). It is real for
$0<x<4$ and purely
imaginary for $x>4$. The combination $H(r,0,x)/\sqrt{x(4-x)}$ is real
on the entire $x>0$ axis. Using consistently the same prescription for
$H(r,0,x)$ and for $\sqrt{x(4-x)}$, we find that the ratio is real and
independent on the prescription used.

Eq.~(\ref{eqsec}) belongs to the Fuchs class, i.e. it has regular singular
points only, eventually including the point at infinity.
In our case, the singularities  on the real axis are located at $ x = 0 \, ,
\: x = 16 $, while also the point at infinity, $ x = \infty$,
is singular, as can be seen replacing the
variable $x$ with $y=1/x$ and studying the equation in $y=0$.

The solution of the homogeneous equation associated to
Eq.~(\ref{eqsec}) can be expressed in terms of complete elliptic
integrals of the first kind, and the particular solution is expressed
as repeated integrations over the elliptic kernel, as it was discussed
in detail in Ref.~\cite{vonManteuffel:2017hms}.
However, in this paper we are going to use another approach for the
solution of the second-order differential equation. We will use the
differential equation to generate power series expansions around the
singular points and at infinity. Each series is determined up to two
arbitrary constants. We will impose the constants of the series in
$x=0$, since we know the initial conditions for ${\mathcal T}_{9}$ in
that point. Then,  the series are matched two-by-two in a point which
lies inside both convergence domains. In this way, we will be able to
fix all the constants and have a representation by series on the whole
real axis.
Our ultimate goal is to be able to evaluate precisely the function
${\mathcal T}_{9}$ on the whole real axis. In order to achieve the
required precision it can be useful to supplement the original
expansion in $x=0$, $x=16$ and infinity, with additional expansions
around regular points.

Once the first master integral ${\mathcal T}_{9}$ has been determined,
we can find the expression of the second, ${\mathcal T}_{10}$, using
Eq.~(\ref{eq1}): \be {\mathcal T}_{10} = \frac{x}{4m^2} \frac{d
  {\mathcal T}_{9}}{dx} + \frac{1}{2m^2} {\mathcal T}_{9} \, .
\label{T10}
\ee

\section{$\tnove$ evaluation for $x \geq 0$}
\label{sec3}
In this Section we discuss the  solution of Eq.~(\ref{eqsec}) in the
region $x \geq 0$. $\tnove$ is obtained  through the series in the singular
regular  points $x=0,\: x=16$ and $x=\infty$  that are then  matched
to cover the entire region $x \geq 0$.   
In all points, we first solve the homogeneous equation
and then the complete equation, 
obtaining all the coefficients of
the series in terms of the first two unknown coefficients.
These unknowns
will be fixed  from
the behaviour of the solution in one point, with the matching procedure.

\subsection{The solution around $x=0$}
\label{sec3.1}

The point $x=0$ allows us to impose the initial conditions and,
therefore, to determine the two constants of integration that come
from the general solution of the second-order linear differential
equation (\ref{eqsec}). For this purpose, it is sufficient to know the
behaviour of the master ${\mathcal T}_{9}$ for $x \to 0$ that can be obtained,
for example, via a large-mass asymptotic expansion of the integral,
 \be
{\mathcal T}_{9} \, \sim \, \log{x} \quad \mbox{for}~ x \to 0.  
\ee
This implies that in the solution no terms with inverse powers of $x$
appear, fixing the constants of integration.

We first consider the homogeneous equation
\be
\frac{d^2 \tnovez}{dx^2} + p (x) \frac{d \tnovez}{dx} + q(x)
     \tnovez = 0 \, .
\label{homog}
\ee
The functions $p(x)$ and $q(x)$ have the following expansion in $x=0$:
\bea
p(x) & \simeq & \frac{4}{x} - \frac{1}{16} - \frac{x}{256} - \frac{x^2}{4096} +
... \, , \\
q(x) & \simeq & \frac{9}{4x^2} - \frac{7}{64x} - \frac{7}{1024} -
\frac{7x}{16384} 
- \frac{7x^2}{262144} + ... \, . 
\eea
Since $x=0$ is a singular regular point, we look for a power series solution
of the form:
\be
\tnovez(x) = x^{\alpha} \sum_{n=0}^{\infty} a_n \, x^n \, ,
\label{solpowser0}
\ee
where $a_n$ are numerical coefficients determined from the differential
equation itself and from the initial conditions.
Substituting the solution (\ref{solpowser0}) in Eq.~(\ref{homog}), we obtain
the characteristic  equation for the determination of $\alpha$:
\be
\left( \alpha+\frac{3}{2} \right)^2 = 0 \, ,
\ee
with double solution in $\alpha=-3/2$. This implies two independent prefactors
of the form $1/(x\sqrt{x})$ and $\log{x}/(x\sqrt{x})$. Therefore, the general
solution of the homogeneous differential equation (\ref{homog}) is
\be
\tnovez(x)  = \frac{1}{\sqrt{x}} \sum_{n=-1}^{\infty} a_n  x^n 
 +  \frac{\log{x}}{\sqrt{x}}  \sum_{n=-1}^{\infty} b_n  x^n \, , 
\label{t9seriesx0}
\ee
where we have absorbed a $1/x$ factor inside the series. The
series (\ref{t9seriesx0}) converges in a circle of radius $r=16$,
i.e. up to the nearest divergence point on the real axis.

Expanding now the differential equation (\ref{homog}) in $x=0$ and
substituting the general solution (\ref{t9seriesx0}), we can fix all
the coefficients of the series in terms of the first two coefficients,
$a_{-1}$ and $b_{-1}$. The first few coefficients are:
\begin{align}
a_0 & = \frac{1}{64} a_{-1} + \frac{1}{32} b_{-1} \, , 
& b_0 & = \frac{1}{64} b_{-1} \, , \\
a_1 & = \frac{9}{16384} a_{-1} + \frac{21}{16384} b_{-1} \, ,
& b_1 & = \frac{9}{16384} b_{-1} \, , \\
a_2 & = \frac{25}{1048576} a_{-1} + \frac{185}{3145728} b_{-1} \, , 
& b_2 & = \frac{25}{1048576} b_{-1} \, .
\end{align}
The general solution for $\tnovez$ 
is a combination of two independent solutions, that can be found imposing,
for instance, $a_{-1}=0$ and $b_{-1}=1$ or $a_{-1}=1$ and $b_{-1}=0$.

Let us now consider Eq.~(\ref{eqsec}). The expansion
of the function $\Omega(x)$ around $x=0$ is\footnote{In order to
  simplify the notation from now on we set $m^2=1$.}
\be
\Omega(x) =
\sum_{n=-2}^{\infty} k_n x^n + \log x \sum_{n=-2}^{\infty} r_n x^n \,
, \ee with first coefficients
\begin{align}
k_{-2} & =  \frac{1}{128}  \, ,
&
r_{-2} & = - \frac{3}{128} \, ,
\\
k_{-1} & =  \frac{21}{2048}  \, ,
&
r_{-1} & = - \frac{11}{2048} \, ,
\\
k_{0} & =  \frac{10549}{7372800}  \, ,
&
r_{0} & = - \frac{183}{163840} \, ,
\end{align}
Therefore, the inhomogenous term has a double pole in $x = 0$, multiplied
also by a single $\log{x}$.
We look for a particular solution of Eq.~(\ref{eqsec}) in $x=0$ of the form:
\be
\widetilde{\tnove}(x)  =  
\sum_{n=-1}^{\infty} p_n  x^n
 +  \log x  \sum_{n=-1}^{\infty} q_n  x^n \, .
\label{parsol0}
\ee
Substituting Eq.~(\ref{parsol0}) in the second-order differential
equation expanded around $x=0$ we obtain, as in the case of the
general solution of the homogeneous equation, terms $p_n$ and $q_n$
that depend on $p_{-1}$ and $q_{-1}$. However, in this case we are
looking for a particular solution, since the general solution of the
homogeneous equation is already known by Eq.~(\ref{t9seriesx0}). We
can then choose to set
\be p_{-1} = 0 \, , \quad q_{-1} = 0 \, ,
\ee
finding the following first terms of the series in
Eq.~(\ref{parsol0}):
\begin{align}
p_0 & = \frac{5}{288} \, ,  \label{pippo}
&
q_0 &= - \frac{1}{96} \, ,
\\
p_1 & = \frac{77}{28800} \, ,
&
q_1 &= - \frac{1}{960} \, , \label{pluto} 
\\
p_2 & = \frac{1237}{5644800} \, ,
&
q_2 &= - \frac{1}{8960} \, .
\label{coeffx0part}
\end{align}
The general solution of the complete equation is therefore:
\be
\tnove(x)  = \frac{1}{\sqrt{x}} \sum_{n=-1}^{\infty} a_n  x^n 
 +  \frac{\log{x}}{\sqrt{x}}  \sum_{n=-1}^{\infty} b_n  x^n 
 + \sum_{n=0}^{\infty} p_n  x^n
 +  \log x  \sum_{n=0}^{\infty} q_n  x^n \, .
\ee
To determine completely the solution, we have to impose the initial conditions. 
Since $\tnove(x)$ can have at most a logarithmic singularity for $x \to 0$,
the coefficients of the power singularities must vanish:
\be
a_{-1}  =  0 \, , \quad b_{-1}  =  0 \, ,
\ee
and, as a consequence, all the $a_n$ and $b_n$ coefficients vanish.

Therefore, the solution of the complete equation reduces to
\be
\tnove(x)
= \sum_{n=0}^{\infty} p_n  x^n
 + \log x   \sum_{n=0}^{\infty} q_n  x^n \, ,
\label{Eq0}
\ee
where the first few coefficients $p_n$ and $q_n$ are given in
Eqs.~(\ref{pippo}--\ref{coeffx0part}).

The solution given in Eq.~(\ref{Eq0}) is real for $x > 0$. However, in
the physical region, $x < 0$ ($s>0$), it develops an imaginary part
that can be determined using the Feynman prescription $x \to -s-i0^+$. 
This means that the logarithmic terms develop an explicit
imaginary part: \be \log x \to \log s - i \pi \, .
\label{analitcontlog}
\ee
Then, $\tnove(x)$ becomes complex for $x < 0$ ($s>0$) with:
\bea
{\rm Re} \, \tnove(s) & = & \sum_{n=0}^{\infty} p_n  (-s)^n 
+ \log s \sum_{n=0}^{\infty} q_n  (-s)^n \, ,
\\
{\rm Im} \, \tnove(s) & = & - \pi \, 
\sum_{n=0}^{\infty} q_n \, (-s)^n \, .
\eea

\subsection{The solution around $x \, = \, 16$}
\label{sec3.2}

The series in $x=0$ is completely determined. The following singular
regular point we have to consider is $x=16$. Since the singular point
closest to $x=16$ is $x=0$, the radius of convergence of the series in
$x=16$ is $r=16$.

By solving the characteristic equation as in the previous subsection, we
obtain a double zero in zero, so that the homogeneous equation has 
a solution of the form:
\be
\tnovez(x)  =  \sum_{n=0}^{\infty} a_n  (x - 16)^n
 +  \log (x - 16)  \sum_{n=0}^{\infty} b_n  (x - 16)^n \, .
\ee
The coefficients are, of course, different from the ones of the
previous section, although we use the same notation to avoid introducing
too many symbols.
The first few coefficients read:
\begin{align}
a_1 & = - \frac{7}{64} a_0
          - \frac{1}{32} b_0 \, ,
&
b_1 & = - \frac{7}{64} b_0 \, , 
\\
a_2 & = \frac{153}{16384} a_0
          + \frac{69}{16384} b_0 \, ,
&
b_2 & = \frac{153}{16384} b_0 \, ,
\\
a_3 & =  - \frac{759}{1048576} a_0
          - \frac{1283}{3145728} b_0 \, ,
&
b_3 & = - \frac{759}{1048576} b_0 \, .
\end{align}

The expansion of the inhomogenous term around $x=16$ is of the following form:
\be
\Omega(x) = \sum_{n=-1}^{\infty} q_n (x - 16)^n \, ,
\label{known16}
\ee
where the first three coefficients $q_n$ are:
\bea
q_{-1} & = & 
          - \frac{3}{4096  \sqrt{3}} \mbox{Li}_2(-7+4 \sqrt{3})
          - \frac{3}{16384  \sqrt{3}} \log^2(7-4 \sqrt{3})
          + \frac{5}{8192}  \log^2(2+\sqrt{5}) \nn\\
&&          - \frac{3}{8192 \sqrt{3}}  \zeta_2 \, ,
\\
q_{0} & = & 
            \frac{19}{131072  \sqrt{3}} \mbox{Li}_2(-7+4 \sqrt{3})
          + \frac{19}{524288  \sqrt{3}} \log^2(7-4 \sqrt{3})
         - \frac{15}{131072}  \log^2(2+\sqrt{5}) \nn\\
&&        + \frac{5}{65536  \sqrt{5}} \log(2+\sqrt{5}) 
          - \frac{1}{16384}  \log(2)
          + \frac{19}{262144 \sqrt{3}} \zeta_2 \, ,
\\
q_{1} & = & 
          - \frac{161}{8388608  \sqrt{3}} \mbox{Li}_2(-7+4 \sqrt{3})
          - \frac{161}{33554432  \sqrt{3}} \log^2(7-4 \sqrt{3})   
	  + \frac{15}{1048576}  \log^2(2+\sqrt{5}) \nn\\
&&       
          - \frac{69}{4194304  \sqrt{5}} \log(2+\sqrt{5}) 
	  + \frac{15}{1048576}  \log(2)
          - \frac{161}{16777216 \sqrt{3}} \zeta_2 \, .
\eea
In particular, note that there is no logarithmic term in Eq.~(\ref{known16}).

The particular solution of the non-homogeneous equation in $x=16$ reads:
\be
\widetilde{\tnove}(x)  \, = \, \sum_{n=0}^{\infty} r_n  (x - 16)^n
 +  \log (x - 16)  \sum_{n=0}^{\infty} p_n  (x - 16)^n \, .
 \label{partx16}
 \ee
 The coefficients $r_i$ and $p_i$ depend on the $r_0$ and $p_0$,
which are undetermined.  However, since we are looking for a
particular solution we can set from the beginning $r_0=0$ and
$p_0=0$. This, in turn, forces the entire series of the logarithmic
part of Eq.~(\ref{partx16}) to vanish,
$ p_n=0 \quad \mbox{for  all}\quad n=1,2,...$ .
Therefore, we have a simple
power series, with the first three terms given by:
\bea r_{1} & = & -
\frac{3}{4096 \sqrt{3}} \mbox{Li}_2(-7+4 \sqrt{3}) - \frac{3}{16384
  \sqrt{3}} \log^2(7-4 \sqrt{3}) + \frac{5}{8192} \log^2(2+\sqrt{5})
\nn\\ && - \frac{3}{8192 \sqrt{3}} \zeta_2 \, , \\ 
r_{2} & = &
\frac{107}{1048576 \sqrt{3}} \mbox{Li}_2(-7+4 \sqrt{3}) +
\frac{107}{4194304 \sqrt{3}} \log^2(7-4 \sqrt{3}) + \frac{5}{262144
  \sqrt{5}} \log(2+\sqrt{5}) \nn\\ && - \frac{175}{2097152}
\log^2(2+\sqrt{5}) - \frac{1}{65536} \log(2) + \frac{107}{2097152
  \sqrt{3}} \zeta_2 \, , \\ 
r_{3} & = & - \frac{6133}{603979776
  \sqrt{3}} \mbox{Li}_2(-7+4 \sqrt{3}) - \frac{6133}{2415919104
  \sqrt{3}} \log^2(7-4 \sqrt{3}) \nn\\ 
&& - \frac{157}{50331648
  \sqrt{5}} \log(2+\sqrt{5}) + \frac{9865}{120795955}
\log^2(2+\sqrt{5}) + \frac{11}{4194304} \log(2) \nn\\ && -
\frac{6133}{1207959552 \sqrt{3}} \zeta_2 \, .
\eea

The general solution of the differential equation is given by
\be
\tnove(x)  =  \sum_{n=0}^{\infty} a_n  (x - 16)^n
 +  \log (x - 16)  \sum_{n=0}^{\infty} b_n  (x - 16)^n 
 + \sum_{n=0}^{\infty} r_n  (x - 16)^n \, .
 \label{x16complete}
\ee
Note that the integral $\tnove(x)$ should be real in the Euclidean
region. However, the logarithmic terms, that come from the
homogeneous solution, are responsible of the appearance of an
imaginary part that cannot be there. We have, therefore, to impose
that $b_0=0$. This condition implies that the logarithmic part of the
expansion vanishes completely. The solution in Eq.~(\ref{x16complete})
becomes a simple power series and depends on a single condition,
$a_0$, that can be fixed as explained in the following section.

\subsection{Matching the series in $x=0$ and $x=16$}
\label{sec3.3}

The series expansion around $x=0$ is completely determined by imposing
the initial conditions.  The series in $x=16$, instead, depends on a
single undetermined constant of integration, $a_0$.  We can compute
$a_0$, imposing that the series in $x=0$ and the one in $x=16$ assume
the same value in a given point in the intersection of the respective
domains of convergence.  Since both series have radius of convergence
$r=16$, in principle it is sufficient to impose that both series have
the same value in any point $x \in (0,16)$.

Dealing with infinite series would exactly determine the coefficient
$a_0$. However, we can only determine an arbitrary, but finite, number
of coefficients of both series. Therefore, $a_0$ will be determined in
an approximate way.

The number of terms in the series depend on the relative precision at
which we want to be able to compute $\tnove(x)$ in a
given point of the real axis. Our goal is to provide a double precision 
numerical routine,
using a relatively small number of terms in the series
(around 50 or less).

If we want to use just the series in $x=0$ and $x=16$ and we want to
be able to provide such a precision, we have to deal with a large number
of terms in the series. In order to keep the number of terms of the order of
50, and relative double precision within the interval
$0<x<16$, we have to add series in intermediate points.

All the points in the interval $x \in (0,16)$ are regular points for the
differential equation and they will result in simple power series
(without the logarithmic part).
In particular, we added series in $x=2$, 4, and 8. The procedure of
matching is, therefore, performed as follows. We match the series in
$x=0$ with the one in $x=2$. As a matching point we choose
$x=1.5$. Then, in the point $x=3.25$ the series in $x=2$ is matched with the one in
$x=4$, while in $x=6$, the  series in $x=4$ is matched with the one in
$x=8$ . Finally, the series in $x=8$ is matched with the one
in $x=16$, in the point $x=12$.

The actual point in which we match two series is of course
arbitrary. Nevertheless, a bad choice would lower the precision of the
matched series.    This would, in
turn, lower the precision for all $x$ above the matching point.  A
possible approach for a good choice is the following. We first start
with a point that assures a good convergence for both the series and
we determine the unknown constants. Then, we vary a bit the point of
the matching and we look at the corresponding variation of the
significant digits of the constants. A good matching point maximises
the number of stable digits in the result for the unknown constants.

\subsection{The solution around $x \, = \, \infty$ }
\label{sec3.4}
We consider now the expansion of $\tnove$ around 
$x \, = \, \infty$. Since the closest singularity to $x=\infty$ is at $x=16$,
we expect the expansion around infinity to be convergent outside the circle of 
radius 16, i.e. for $|x|>16$.

The expansion at infinity can be studied systematically by performing
the following change of variable $ x = 1/y $ and, then,
considering the limit $ y \to 0 $. 

The homogeneous equation in $y \to 0$ limit reads
\be
\frac{d^2  \tnovez}{dy^2} - \frac{3}{y} \frac{d\tnovez}{dy} + \frac{4}{y^2}
\tnovez = 0 \, .
\ee
We look for a solution of the form
\be
\tnovez(y) = y^{\beta} \sum_{n=0}^{\infty} A_n y^n.
\ee
The characteristic equation gives $(\beta-2)^2 = 0$, with a double zero in
$\beta = 2$. Therefore, the solution of the homogeneous equation, in the
original variable $x=1/y$ is
\be
\tnovez(x) = \sum_{n=2}^{\infty}
\frac{a_n}{x^n} - \log x \sum_{n=2}^{\infty} \frac{b_n}{x^n} \, ,
\ee
with the coefficients $a_n$ and $b_n$ expressed in terms of the
lowest-order ones, $a_2$ and $b_2$ as shown for the first few terms:
\begin{align}
a_3 &= 4 \, a_2  +  8 \, b_2 \, ,
&
b_3 &= 4 \, b_2 \, ,
\\
a_4 &=  36 \, a_2  +  84 \, b_2 \, ,
&
b_4 &= 36 \, b_2 \, ,
\\
a_5 &=  400 \, a_2  + 2960 \, b_2 \, ,
&
b_5 &= 400 \, b_2 \,  .
\end{align}

The expansion of the non-homogeneous term $\Omega(x)$ around $x=\infty$ is of the form:
\be
\Omega(x) =  \sum_{n=0}^{\infty} \frac{k_n}{x^n}
 -  \log x  \sum_{n=0}^{\infty} \frac{l_n}{x^n}
 +  \log^2 x  \sum_{n=0}^{\infty} \frac{m_n}{x^n} \, ,
\ee
where the lowest-order coefficients read:
\begin{align}
k_0 &= - \frac{3}{4} \zeta_2  \, ,
&
l_0 &= 0 \, ,
&
m_0 &=   \frac{1}{4} \, ,
\\
k_1 &=  \frac{3}{2}  -  \frac{27}{2} \zeta_2 \, ,
&
l_1 &= - 4 \, ,
&
m_1&  =  \frac{13}{4} \, ,
\\
k_2 &=  \frac{245}{8}  -  \frac{441}{2} \zeta_2 \, ,
&
l_2 &= - \frac{131}{2} \, ,
&
m_2 &=   \frac{199}{4} \,  .
\end{align}

The differential equation involves a second derivative and the non-homogeneous term has 
double logarithmic terms. Therefore, the particular solution must contain up to 
four powers of the logarithm:
\be
\widetilde{\tnove}(x) =  \sum_{n=2}^{\infty} \frac{p_n}{x^n}
 -  \log x  \sum_{n=2}^{\infty} \frac{q_n}{x^n}
 +  \log^2 x  \sum_{n=2}^{\infty} \frac{r_n}{x^n}
 -  \log^3 x  \sum_{n=2}^{\infty} \frac{u_n}{x^n}
 +  \log^4 x  \sum_{n=2}^{\infty} \frac{t_n}{x^n} \, .
\label{partxoo}
\ee
Substituting Eq.~(\ref{partxoo}) into the non-homogeneous equation, we
obtain the following first few coefficients: 
\begin{align}
p_2 &= 0 \, , & p_3 &= 7 + \frac{3}{2} \zeta_2 \, , & p_4 &= \frac{1075}{16} - \frac{15}{8} \zeta_2  \, ,
\\
q_2 &= 0 \, , & q_3 &= - 1 - 6 \zeta_2 \, , & q_4 &= - \frac{91}{4} - 63 \zeta_2  \, ,
\\
r_2 &= -  \frac{3}{8} \zeta_2 \, , & r_3 &= - \frac{7}{4} - \frac{3}{2} \zeta_2 \, , & r_4 &=  - \frac{185}{16}
- \frac{27}{2} \zeta_2 \, ,
\\
u_2 &= 0 \, , & u_3 &= \frac{2}{3} \, , & u_4 &= 7   \, ,
\\
t_2 &= \frac{1}{48} \, , & t_3 &= \frac{1}{12} \, ,  & t_4 &= \frac{3}{4} \, .
\label{coeffsxoo}
\end{align}

Finally, the general solution is given by:
\be
\tnove(x) = \sum_{n=2}^{\infty} \frac{\tilde{p}_n}{x^n}
 -  \log x  \sum_{n=2}^{\infty} \frac{\tilde{q}_n}{x^n}
 +  \log^2 x  \sum_{n=2}^{\infty} \frac{r_n}{x^n}
 -  \log^3 x  \sum_{n=2}^{\infty} \frac{u_n}{x^n}
 +  \log^4 x  \sum_{n=2}^{\infty} \frac{t_n}{x^n},
\label{Eqxoo}
\ee
where we set: 
\be
\tilde{p}_n \, = \, p_n + a_n \, , \quad
\tilde{q}_n \, = \, q_n + b_n \, .
\ee
The coefficients of the power series and of the single logarithm depend upon the
two constants of integration, while the coefficients of the double, triple
and fourth logarithm are uniquely determined.
As in the case of the series in $x=16$, the two constants have to be
determined matching the solution in $x=\infty$ with the one in $x=16$,
in an intermediate point chosen in the range $16<x<32$ (the series in
$x=16$ has radius of convergence $r=16$).
However, in order to improve the precision in the determination of the
integral $\tnove(x)$, without adding too many terms in the series
expansions, it is better to add the expansions in three additional
regular points: $x=32$, $x=64$ and $x=128$, before the matching with
$x=\infty$.

\section{$\tnove$ evaluation for $x < 0$ ($s > 0$)}
\label{sec4}
The solution for $\tnove$ in the region $x < 0$ can be constructed
starting from
the expansion of the amplitude $\tnove(x)$ for large time-like momenta,
namely for $x \to -\infty$ ($s \to \infty$), that can be found from the
asymptotic expansion in the space-like region ($x \to \infty$) by analytic
continuation. 
With the Feynman prescription
\be
x \, \to \, - s - i 0^+  \, ,
\ee
we have to consider that the logarithm develops an immaginary part as in
Eq.~(\ref{analitcontlog}):
\be
\log x  \to  \log s  -i \pi \, .
\ee
Then $\tnove(s)$  becomes complex and its real and immaginary parts are
given by:
\bea
{\rm Re} \, \tnove(s)  & = &   \sum_{n=2}^{\infty} (-1)^n \frac{\tilde{p}_n}{s^n}
  -   \log s  \sum_{n=2}^{\infty} (-1)^n \frac{\tilde{q}_n}{s^n}
  +  (\log^2  s  -  \pi^2)  \sum_{n=2}^{\infty} (-1)^n \frac{r_n}{s^n} - \bigl( \log^3  s  \nn\\
& &  -  3 \pi^2 \log s \bigr)  \sum_{n=2}^{\infty}
 (-1)^n \frac{u_n}{s^n}  +  \left( \log^4 s - 6 \pi^2 \log^2 s + \pi^4 \right)  
\sum_{n=2}^{\infty} (-1)^n \frac{t_n}{s^n} \, , \\
{\rm Im} \, \tnove(s)  & = &   \pi 
\Bigg[ 
 - \sum_{n=2}^{\infty} (-1)^n \frac{\tilde{q}_n}{s^n}
 + 2 \log s  \sum_{n=2}^{\infty} (-1)^n \frac{r_n}{s^n}
 - \left( 3 \log^2 s - \pi^2 \right)  \sum_{n=2}^{\infty} (-1)^n \frac{u_n}{s^n} \nn\\
  & &   + \left( 4 \log^3 s - 4 \pi^2 \log s \right) \sum_{n=2}^{\infty} (-1)^n 
 \frac{t_n}{s^n}
\Bigg] \, .
\eea 

The series in $x=0$ has a convergence radius $r=16$ and the series at
infinity converges in $|x|>16$. In order to determine $\tnove$ in
all points of the time-like region with the required precision, we need
additional expansion points to sew
the series at infinity with the one in $x=0$. Since in the region
$-\infty < x<0$ ($0<s<\infty$) there are no singular points, the points to be
added will be regular points, and the corresponding series will be simple power
series.

We added the following points: $s=4$, $s=8$, $s=16$,
$s=32$, $s=64$ and finally $s=128$. We will discuss extensively just $s=16$.

\subsection{The solution around $s = 16$ }
\label{sec4.1}

The point $s=16$  is a regular point. Therefore, the expansion of the
homogeneous solution is a power series
\be
{\mathcal T}_9^{(0)}(s) = \sum_{n=0}^{\infty} a_n (s-16)^n \, ,
\ee
with the first few coefficients given in terms of $a_0$ and $a_1$ by
\bea
a_2 & = & - \frac{25}{4096} a_0
          - \frac{9}{64} a_1 \, , \\
a_3 & = & \frac{53}{65536} a_0
          + \frac{57}{4096} a_1 \, , \\
a_4 & = & - \frac{7859}{100663296} a_0
          - \frac{39}{32768} a_1 \, . 
\eea

The expansion of the inhomogneous term around $s=16$ is of the following form
\be
\Omega(s) = \sum_{n=1}^{\infty} q_n (s-16)^n \, ,
\ee
where
\bea
q_1 & = & 
            \frac{5}{2097152 \sqrt{3}} \log{(2+\sqrt{3})}
          + \frac{51}{10485760 \sqrt{5}} \mbox{Li}_2 \left( \frac{1}{(2+\sqrt{5})^2} \right) \nn\\
& &          + \frac{51}{10485760 \sqrt{5}} \log(2+\sqrt{5})^2
          - \frac{35}{8388608}  \log{(2+\sqrt{3})}^2
          - \frac{3}{2621440}  \log{(2)} \nn\\
& &
          - \frac{51}{10485760 \sqrt{5}}  \zeta(2)
          + \frac{105}{16777216}  \zeta(2)
          - i \pi \biggl[ \frac{5}{4194304 \sqrt{3}}
            + \frac{51}{10485760 \sqrt{5}} \log(2+\sqrt{5})  \nn\\
& &
          - \frac{3}{10485760} 
          - \frac{35}{8388608} \log{(2+\sqrt{3})} \biggr] \, , \\
q_2 & = & 
          - \frac{245}{402653184 \sqrt{3}} \log{(2+\sqrt{3})}
          - \frac{4389}{6710886400 \sqrt{5}} \mbox{Li}_2 \left( \frac{1}{(2+\sqrt{5})^2} \right) \nn\\
& &
          - \frac{4389}{6710886400 \sqrt{5}} \log(2+\sqrt{5})^2
          + \frac{1}{62914560} 
          + \frac{155}{268435456}  \log{(2+\sqrt{3})}^2 \nn\\
& &
          + \frac{231}{838860800}  \log{(2)}
          + \frac{4389}{6710886400 \sqrt{5}}  \zeta(2)
          - \frac{465}{536870912}  \zeta(2)
	  - i \pi \biggl[
          - \frac{245}{805306368 \sqrt{3}}  \nn\\
& &
          - \frac{4389}{6710886400 \sqrt{5}}  \log(2+\sqrt{5})
          + \frac{231}{3355443200} 
          + \frac{155}{268435456} \log{(2+\sqrt{3})} \biggr] \, , \\
q_3 & = & 
            \frac{205}{2147483648 \sqrt{3}} \log{(2+\sqrt{3})}
          + \frac{3819}{53687091200 \sqrt{5}} \mbox{Li}_2 \left( \frac{1}{(2+\sqrt{5})^2} \right) \nn\\
& &
          + \frac{3819}{53687091200 \sqrt{5}} \log(2+\sqrt{5})^2
          - \frac{233}{48318382080} 
          - \frac{555}{8589934592}  \log{(2+\sqrt{3})}^2 \nn\\
& &
          - \frac{547}{13421772800}  \log{(2)}
          - \frac{3819}{53687091200 \sqrt{5}}  \zeta(2)
          + \frac{1665}{17179869184}  \zeta(2) \nn\\
& &
          - i \pi \biggl[ \frac{205}{4294967296 \sqrt{3}}
          + \frac{3819}{53687091200 \sqrt{5}} \log(2+\sqrt{5})
          - \frac{547}{53687091200}  \nn\\
& &
          - \frac{555}{8589934592} \log{(2+\sqrt{3})} \biggr] \, .
\eea
Therefore, the particular solution of the differential equation in
$s=16$ is, again, a power series in which the coefficients $p_n$, $n
\geq 2$, depend upon the first two coefficients, $p_0$ and $p_1$.
Since we are now looking for a particular solution, we can set $p_0=0$
and $p_1=0$, finding
\be
\widetilde{\tnove}(s) = \sum_{n=2}^{\infty} p_n (s-16)^n \, ,
\ee
with the first few coefficients that read
\bea
p_2 & = & 
       - \frac{3}{262144 \sqrt{5}} \mbox{Li}_2 \left( \frac{1}{(2+\sqrt{5})^2} \right)
       - \frac{3}{262144 \sqrt{5}} \log(2+\sqrt{5})^2
       + \frac{5}{524288} \log(2+\sqrt{3})^2 \nn\\
& &       + \frac{3}{262144 \sqrt{5}} \zeta(2)
       - \frac{15}{1048576} \zeta(2)
       + i \pi \biggl[ \frac{3}{262144 \sqrt{5}}  \log(2+\sqrt{5}) \nn\\
& & 
       - \frac{5}{524288} \log(2+\sqrt{3}) \biggr] \, , \\
p_3 & = & 
       + \frac{5}{12582912 \sqrt{3}} \log(2+\sqrt{3})
       + \frac{79}{41943040 \sqrt{5}} \mbox{Li}_2 \left( \frac{1}{(2+\sqrt{5})^2} \right) \nn\\
& & 
       + \frac{79}{41943040 \sqrt{5}} \log(2+\sqrt{5})^2
       - \frac{5}{3145728} \log(2+\sqrt{3})^2
       - \frac{1}{5242880} \log{(2)} \nn\\
& & 
       - \frac{79}{41943040 \sqrt{5}} \zeta(2)
       + \frac{5}{2097152} \zeta(2)
       - i \pi \biggl[
         \frac{5}{25165824 \sqrt{3}} 
       + \frac{79}{41943040 \sqrt{5}} \log(2+\sqrt{5}) \nn\\
& & 
       - \frac{1}{20971520} 
       - \frac{5}{3145728} \log(2+\sqrt{3}) \biggr] \, , \\ 
p_4 & = & 
       - \frac{95}{1207959552 \sqrt{3}} \log(2+\sqrt{3})
       - \frac{11111}{53687091200 \sqrt{5}} \mbox{Li}_2 \left( \frac{1}{(2+\sqrt{5})^2} \right) \nn\\
& & 
       - \frac{11111}{53687091200 \sqrt{5}} \log(2+\sqrt{5})^2
       + \frac{1}{754974720}
       + \frac{2275}{12884901888} \log(2+\sqrt{3})^2 \nn\\
& & 
       + \frac{61}{1677721600} \log{(2)}
       + \frac{11111}{53687091200 \sqrt{5}} \zeta(2)
       - \frac{2275}{8589934592} \zeta(2) \nn\\
& & 
       + i \pi \biggl[ \frac{95}{2415919104 \sqrt{3}} 
       + \frac{11111}{53687091200 \sqrt{5}} \log(2+\sqrt{5})
       - \frac{61}{6710886400}  \nn\\
& & 
       - \frac{2275}{12884901888} \log(2+\sqrt{3}) \biggr] \, .
\eea

Finally, the general solution is given by
\be
\tnove(s) = \sum_{n=0}^{\infty} w_n (s-16)^n \, ,
\ee
where
\be
w_0 = a_0 \, , \quad w_1 = a_1 \, , \quad w_n = a_n+p_n \; \; \mbox{for} \; n \geq 2 \, .
\ee

\section{Expansions for the Master Integral ${\mathcal T}_{10}$ \label{SecMI}}

The second MI is directly determined from the first one by means of
Eq.~(\ref{T10}):
\be
   {\mathcal T}_{10} = \frac{x}{4m^2} \frac{d {\mathcal T}_{9}}{dx}  +
   \frac{1}{2m^2} 
{\mathcal T}_{9} \, .
\label{T10bis}
\ee
Knowing the series expressions for ${\mathcal T}_{9}$,
Eq.~(\ref{T10bis}) allows to determine ${\mathcal T}_{10}$ performing
a simple derivative.

The matching conditions that we imposed for the series expansions in
the various points of the real axis, for the complete determination of
${\mathcal T}_{9}$, are still valid for ${\mathcal T}_{10}$. In principle, 
one simply has to take the derivative of each of these series, the series 
itself, and combine them in order to fulfill Eq.~(\ref{T10bis}). 
In the case of infinite series, there would be no
difference in the determination and precision of ${\mathcal T}_{10}$
with respect to what we found for ${\mathcal T}_{9}$. However, we deal
with truncated series and this means that the optimal choice for a
matching point of two series for ${\mathcal T}_{9}$ can be less
optimal for the corresponding series of ${\mathcal T}_{10}$. 
Therefore, we decided to determine the matching points independently for
the series of ${\mathcal T}_{9}$ and ${\mathcal T}_{10}$.

We used the criterium of maximization of the number of stable digits 
in the determination of the unknown constants, under the variation of the 
matching point. In so doing, we found that the matching points for 
corresponding pair of series of ${\mathcal T}_{9}$ and ${\mathcal T}_{10}$ 
give rise to slightly different matching constants.
We used the difference between the values of the matching constants as an
indicator for the precision at which we can claim the series reproduce the
numerical value of the masters. In all the matching points, we found
corresponding matching constants 
that agree with double precision.

\section{The Fortran Routine }
\label{fortranroutine}

In this section we give details on the numerical routines that accompany  the
paper.

The routine implements the series in the various points of the real domain 
discussed in the previous sections. In some points (in particular in 
$x=2,4,8,16,32,64,128$ and $s=4,8,12,16,32,64,128$), in order to improve 
the convergence of the series, we performed the Bernoulli variable
transformation
\cite{tHooft:1978jhc, Maximon2807}, which is defined as
\begin{equation}
t=\log\left( \frac{b-x_0}{x_0-a}\frac{x-a}{b-x}\right) \label{eq:bernoulli}
\end{equation}
for a series expansion around $x_0$, with nearest singular points $a$
and $b$, such that $a< x_0 < b$. This change of variable usually
increases the convergence of the series near the point of expansion
(see for instance Refs.~\cite{Pozzorini:2005ff,Aglietti:2007as}).  Althought
in the points indicated above we found a considerable increase in such
convergence, resulting in an increase of the number of reliable digits
of the final result, in $x=0$ and $x=\pm \infty$ the original power
series worked at the same level of accuracy (or sometimes even better)
or had better numerical behaviour. Therefore, the routines are written
using the original series in $x=0$ and $x=\pm \infty$, and the series
in the Bernoulli variable in all the regular points and $x=16$.

The numerical program consists of the header file
\verb|main_elliptic.f| and the two main files \verb|MI1.f| and
\verb|MI2.f|, which compute the master integrals $\mathcal{T}_9$ and
$\mathcal{T}_{10}$ respectively. The program is written in {\tt
  FORTRAN}. Several files contain the lengthy formulae of the
expansions around the various points.

The program can be used in the two following ways:
\begin{itemize}
\item Way 1: As a whole with output onto the screen and into an outputfile.
  After unzipping the files
\begin{equation*}
  {\it \text{\it tar -xvf }\text{\it elliptic.zip} \hspace*{0.5cm}
    \text{ or alternatively } \hspace*{0.5cm} \text{\it unzip elliptic.zip}}
\end{equation*}
the program can then be compiled with the provided {\tt makefile}, meaning by
typing
\begin{equation*}
{\it make}
\end{equation*}
 and run by the command
\begin{equation*}
{\it ./run}\; \#\text{value of }x\; \#\text{name of outputfile}
\end{equation*}
If no input value for $x=-s$ is given, the program interrupts and
asks to input a value. If instead no input for the name of the output
file is given, the output is written into a default file named
\verb|output_MI.dat|.
\item Way 2: Inside another program. In this case only the files
  \verb|MI1.f| for $\mathcal{T}_9$ and \verb|MI2.f| for
  $\mathcal{T}_{10}$ are needed as well as all files in the folder
  \verb|seriesexpansions|. The {\tt makefile} of the other program
  must then be adjusted by adding \verb|MI1.o| or \verb|MI2.o| to the
  files to be compiled.  The function {\tt complex*16 MI1(double
    precision x)} for $\mathcal{T}_9$ or {\tt complex*16 MI2(double
    precision x)} for $\mathcal{T}_{10}$ can then be called directly
  within any other {\tt FORTRAN} program.
\end{itemize}

In the following we list the various files and explain them in more details.
\begin{itemize}
\item \verb|main_elliptic.f|: The main file calls the functions MI1(x)
  and MI2(x) for the value $x$ given by the user as an argument when
  running the program and writes the output. This file is not needed
  if the user wants to call the integrals from within his/her own
  program.
\item \verb|MI1.f|: Computes the master integral $\mathcal{T}_9$ and
  can be called by the user directly if he/she decides to call the
  integral from within his/her own program. \verb|MI1.f| decides which
  series expansion is needed for the given value of $x$ and returns
  the corresponding value. It needs the help files that are provided
  in the folder \verb|seriesexpansions|.
\item \verb|MI2.f|: Same as \verb|MI1.f| but for
  $\mathcal{T}_{10}$.
\item \verb|seriesexpansions/MI1_in_x_#.f|: Help files that contain
  the lengthy expressions for the series expansions of $\mathcal{T}_9$
  around $x=0,2,4,8,16,32,64, 128$ and $\infty$, where \verb|#| stands
  for the respective value of $x$.
\item \verb|seriesexpansions/MI1_in_s_#.f|: Help file that contains
  the lengthy expressions for the series expansions of $\mathcal{T}_9$
  around $s=4,8,12,16,32,64,128$ and $\infty$.
\item \verb|seriesexpansions/MI2_in_x_#.f| and
  \verb|seriesexpansions/MI2_in_s_#.f|: Same as
  \verb|seriesexpansions/MI1_in_x_#.f| and
  \verb|seriesexpansions/MI1_in_s_#.f| respectively but for
  $\mathcal{T}_{10}$.
\end{itemize}

\subsection{Numerical Checks}

We performed several numerical checks both to ensure the correctness
of our results and to check the numerical accuracy.  We will
describe them in the following.
\begin{itemize}

\item As outlined in Section~\ref{sec3.3}, we determined the
  constants of integration by matching the series in points within the
  radius of convergence of two series, starting from $x=0$ which we
  have determined completely, going to $x=\infty$ and $x=-\infty$. We
  did this procedure for both $\mathcal{T}_{9}$ and $\mathcal{T}_{10}$
  separately. If we would be able to expand the series to arbitrary
  high order, the constants of integration of $\mathcal{T}_{9}$ and
  $\mathcal{T}_{10}$ would be the same. However, we work with
  truncated series, and the determination of the constants depend upon
  the details of the series used, as for instance the form of the
  coefficients and the number of terms.  As a consequence, the
  integration constants are not exactly the same.  This allows us to
  use the comparison of the matching constants between
  $\mathcal{T}_{9}$ and $\mathcal{T}_{10}$ for each series to
  determine internally the numerical accuracy of the procedure. Doing
  so we find agreement in all the series to double precision accuracy.

\item As an internal check and as a determination of the accuracy of
  the result, we adopted the following strategy. The series in $x=0$
  is completely determined, since we impose the initial conditions in
  that point. Starting from $x=0$, we match the undetermined constants
  of the series as described in the paper up to the series in
  $x=\infty$. Independently, we perform the same procedure in the
  Minkowski region, starting from $s=0$ up to the series in
  $s=\infty$. Now with the series in $x=\infty$, we perform an
  analytic continuation to the Minkowski region, $s>0$, and we
  numerically evaluate the series in $s=1000$, comparing the result
  with the numerical evaluation, in the same point, of the series
  obtained with the matchings in the Minkowski region. We find that
  the two numbers agree 
  with double precision.

\item We cross-checked our numerical routines against PySecDec in
  several points of the entire domain, in both the Euclidean and
  Minkowski regions. We found complete agreement within
  the numerical accuracy of PySecDec, which is limited to 5-6 digits.

\item The most stringent test was the one done with the numbers coming
  from the exact solution of Ref.~\cite{vonManteuffel:2017hms}. We could
  check our routines against the numbers provided by the authors of
  Ref.~\cite{vonManteuffel:2017hms} in $x=3,13,50$ and $s=3,5,18,50$, finding
  an agreement 
  to double precision accuracy\footnote{In the comparison between the numbers of 
  our routines and the numbers of Ref.~\cite{vonManteuffel:2017hms} it must be 
  remembered that the normalization of the integrals are different in the two
  works, as can be seen from our Eq.~(\ref{normalization}) and Eq.~(2.2) of
  Ref.\cite{vonManteuffel:2017hms}.
  In particular, in order to match the numbers of
  Ref.~\cite{vonManteuffel:2017hms}, our $\mathcal{T}_{9}$ has to be multiplied by
  $16$, while $\mathcal{T}_{10}$ for $-16$.}.

\end{itemize}

\section{Conclusions }
\label{conclusions}
In this paper we presented a semi-analytical evaluation for the two MIs of
the crossed vertex topology with a closed massive loop, implemented in a
Fortran numerical routine.

The two MIs can be expressed in power series of the dimensional
parameter $\epsilon = (4-d)/2$. Each order in $\epsilon$ fulfills a
system of two coupled first-order linear differential equations, that
admits solutions in terms of one-fold integrals of complete elliptic
integrals of the first and second kind times polylogarithmic terms
(see Ref.~\cite{vonManteuffel:2017hms}). In the present paper we focus
on the solution of the differential equations for the
${\mathcal  O}(\epsilon^0)$, which is relevant for phenomenological
applications at the NNLO.

In order to implement the solutions in a Fortran numerical routine,
for the precise evaluation of the two MIs, we followed a standard
approach that was used in the past for the study of the equal-mass
sunrise and the three-point function with two massive exchanges,
namely the solution by series of the equivalent second-order linear
differential equation for one of the masters. The other master is then
calculated by a simple derivative, once the first master is known.

Expanding the master in series in the singular points of the
differential equation we were able to directly construct a solution
that covers the entire range $ -\infty \leq x \leq \infty$
which is suitable for precise numerical evaluations.

The Fortran routine presented in this work returns the numerical value
of the two MIs for every real value of the dimensionless parameter
they depend on, with double precision accuracy.

\vspace*{6mm}

\noindent {\large{\bf Acknowledgments}}

\vspace*{2mm}

\noindent 
We wish to thank Andreas von Manteuffel and Lorenzo Tancredi for
having provided us with the numeric results of the two master integrals
published in Ref.~\cite{vonManteuffel:2017hms} in additional points of the
phase spase, for cross-checks.
P.P.G. acknowledges the support of the Spanish Agencia Estatal de Investigacion 
through the grant ``IFT Centro de Excelencia Severo Ochoa SEV-2016-0597". 
R.G. is partially supported by a European Union COFUND/Durham 
Junior Research Fellowship under the EU grant number 609412.
R.G. thanks the Galileo Galilei Institute for Theoretical Physics for the 
hospitality and the INFN for partial support during the completion of this work

\bibliographystyle{utphys}
\bibliography{biblio2}

\end{document}